\documentclass[preprintnumbers,amsmath,amssymb,floatfix,9pt,prd,twocolumn,showpacs,
superscriptaddress,nofootinbib]{revtex4}
\usepackage{graphicx}
\usepackage{epsfig}
\usepackage{bm}
\usepackage{amsfonts}

\begin{document}

\title{Notes on dark energy interacting with dark matter and unparticle
 in loop quantum cosmology}

\author{\textbf{ Mubasher Jamil}} \email{mjamil@camp.nust.edu.pk}
\affiliation{Center for Advanced Mathematics and Physics (CAMP),\\
National University of Sciences and Technology (NUST), H-12,
Islamabad, Pakistan}

\author{\textbf{ D. Momeni}}
\email{d.momeni@yahoo.com}
 \affiliation{
Department of Physics, Faculty of Sciences, Tarbiat Moa'llem
University, Tehran, Iran}

\author{\textbf{  Muneer A. Rashid}}
\email{muneerrshd@yahoo.com}
\affiliation{Center for Advanced Mathematics and Physics (CAMP),\\
National University of Sciences and Technology (NUST),  H-12,
Islamabad, Pakistan}

\begin{abstract}
{\bf Abstract:}  We investigate the  behavior of dark energy
interacting with dark matter and unparticle in the framework of loop
quantum cosmology. In four toy models, we study the interaction
between the cosmic components by choosing different coupling
functions representing the interaction. We found that there are only
two attractor solutions namely dark energy dominated and dark matter
dominated Universe. The other two models are unstable, as they
predict either a dark energy filled Universe or one completely
devoid of it.

\end{abstract}

 \maketitle

\section{Introduction}
Recent cosmological and astrophysical data gathered from the
observations of supernovae SNe Ia {\cite{c1}}, cosmic microwave
background radiations via WMAP {\cite{c2}}, galaxy redshift surveys
via SDSS {\cite{c3}} and galactic X-ray {\cite{c4}} convincingly
suggest that the observable Universe experiences an accelerated
expansion phase. It is well-known that the simplest and elegant way
to explain this behavior is the inclusion of Einstein's cosmological
constant \cite{c7}, however the two deep theoretical problems
(namely the ``fine-tuning'' and the ``coincidence'' one) led to the
notion of `dark energy'. The dynamical nature (i.e. composition and
origin) of dark energy, at least in an effective level, can arise
from various scalar fields, such as a canonical scalar field
(quintessence) \cite{quint}, a phantom field \cite{phant}, that is a
scalar field with a negative sign of the kinetic term, or the
combination of quintessence and phantom in a unified model named
quintom \cite{quintom}. One of the first quintom works which
appeared  is \cite{sergei1}. Recent review on dynamical DE from
modified gravity perspective can be found in \cite{sergei2}. Also,
there are other momentous works on dynamics of the scalar models for
accelerating expansion phase of the universe with different methods
and terminologies\cite{sergei3}. Recently, a new type of the scalar
models have been investigated more than others \cite{sergei4}.

One of the long-standing problems in the standard Big Bang cosmology
is the initial singularity from which all matter and energy
originated. Standard cosmology based on general relativity offers no
resolution to this problem, however a recent quantum gravitational
model of loop quantum gravity (LQG) offers a nice solution. The
theory and principles of LQG when applied in the cosmological
framework creates a new theoretical framework of loop quantum
cosmology (LQC). The effects of loop quantum gravity can be
described in two possible ways: the first one is based on the
modification of the behavior of the inverse scale factor operator
below a critical scale factor. This approach has been used to study
quantum bounces, avoidance of singularities and to produce
inflationary expansion \cite{6}. A second approach is to add a term
quadratic in density to the Friedmann equation. In LQC, the
non-perturbative effects lead to correction term $-\rho^2/\rho_c$ to
the standard Friedmann equation. With the inclusion of this term,
the Universe bounces quantum mechanically as the energy density of
matter-energy reaches the level of $\rho_c$ (order of the Planck
density). Thus the LQC is non-singular by producing a bounce before
the occurrence of any potential singularity and hence transitions
from a pre-Bang and after-Bang are all well-defined. The
observational constraints due to the quadratic term in (1) are
discussed in the literatures \cite{obs} where it is shown that the
model with quadratic correction to density is consistent with the
observational tests. Thus we should not worry on the solar system
tests in this model. For the Universe with a large scale factor, the
first type of modification (first approach) to the effective
Friedmann equation can be neglected and only the second type of
modification (second approach) is important \cite{6}. Thus the
dynamics of dark energy have been investigated recently in LQC using
second approach \cite{de}.

In this paper, we address the problem of cosmic coincidence problem
in a modern theoretical framework of loop quantum cosmology. Here we
assume a non-minimal coupling between dark energy, dark matter and
the unparticle component. Since the nature of dark energy and dark
matter is still unknown, it is possible to have non-gravitational
interactions between them. We are unsure of the form of the
interaction, hence we introduce the interaction terms only
phenomenologically. We construct a system of dynamical equations
containing three equations for the three components. We convert them
to dimensionless form and perform stability analysis. We construct
four toy models and show that the dynamical equations have two
possible attractor solutions i.e. dark matter dominated and dark
energy dominated. Other dynamical systems are unstable i.e. one
model predicts that everything decays and the Universe gets emptier
void of energy. Another model predicts that the Universe will
contain only dark energy.

The model in which dark energy interacts with two different fluids
has been investigated in the literature. In \cite{cruz}, the two
fluids were dark matter and another was unspecified. However, in
another investigation \cite{jamil}, the third component was taken as
radiation to address the cosmic-triple-coincidence problem and study
the generalized second law of thermodynamics. In this article, we
choose the third component as unparticle \cite{georgi}, following
\cite{2} thereby generalizing their study from the general
relativistic cosmology to the loop quantum cosmology. An unparticle
is based on the hypothesis that there could be exact scale invariant
hidden sector resisted at a high energy scale. The fundamental
energy scale of unparticle is far beyond the reach of today's
accelerators, there is a possibility that this new unparticle sector
could affect the low-energy phenomenology. An interesting feature of
unparticle is that it does not have a definite mass and instead has
a continuous spectral density as a result of scale invariance.
Moreover, the equation of state of unparticle $w_u$ is positive
unlike dark energy and it interacts weakly with standard model
particles. We consider the question how the evolution of Universe is
affected when the unparticle takes part in the interaction with dark
energy and dark matter.

This paper is organized as follows: in section II, we construct a
cosmological scenario in which dark energy interact with dark matter
and unparticle. Then we present the formalism of autonomous
dynamical system which is suitable for phase space stability
analysis. In section III, we study the stability of the dynamical
system by choosing different coupling functions. Finally we briefly
discuss our results in the last section.

\section{Our model}

Applying the techniques of loop quantum gravity to homologous and
isotropic spacetime leads to the so-called loop quantum cosmology.
Due to quantum corrections, the Friedmann equations get modified.
The big bang singularity is resolved and replaced by a quantum
bounce \cite{3}. For a brief summary on loop quantum cosmology, see
\cite{4}. Considering quantum correction, the modified Friedmann
equation turns out to be (in the case of $k = 0$) \cite{4}
\begin{equation}\label{1}
H^2=\frac{\kappa^2}{3}\rho\Big(1-\frac{\rho}{\rho_c}\Big),
\end{equation}
where $\rho=\rho_m+\rho_d+\rho_u$, where $\rho_m$, $\rho_d$ and
$\rho_u$ represent the energy densities of matter, dark energy and
the radiation. Also
$\rho_c\equiv\frac{\sqrt{3}}{16\pi^2\gamma^3}\rho_{\text{Pl}}$
determines the loop quantum effects, $\rho_{\text{Pl}}$ is the
Planck density and $\gamma$ is the  dimensionless Barbero$-$Immirzi
parameter. This parameter could be fixed as 0.2375 in order to give
the area formula of black hole entropy in loop quantum gravity
\cite{5}. The observational constraints due to the quadratic term in
(1) are discussed in the literature \cite{obs}, where it is shown
that the model with quadratic correction to density is consistent
with the observational tests. Thus we should not worry on the solar
system tests in this model.

 Due to the corrected term in (\ref{1}), the big bang
singularity is replaced by a quantum bounce happening at $\rho_c$.
The bounce is supposed to happen when the matter$-$energy density
reaches the critical value $\rho_c$. However the numerical
simulations show that modified Friedmann equations are valid in the
whole cosmic evolution including the bounce \cite{3}.

Another FRW equation is
\begin{equation}\label{2}
\dot H=-\frac{\kappa^2}{2}(\rho+p)\Big(1-2\frac{\rho}{\rho_c}\Big).
\end{equation}
For a spatially flat Universe, the total energy conservation
equation is
\begin{equation}\label{1a}
\dot \rho+3H(\rho+p)=0,
\end{equation}
where $H$ is the Hubble parameter, $\rho$ is the total energy
density and $p$ is the total pressure of the background fluid.

We assume a three component fluid containing matter, dark energy and
unparticle having an interaction. The corresponding continuity
equations are \cite{2}
\begin{eqnarray}\label{2a}
\dot \rho_d+3H(\rho_d+p_d)&=&\Gamma_1,\nonumber\\
\dot\rho_m+3H\rho_m&=&\Gamma_2,\\ \dot
\rho_u+3H(\rho_u+p_u)&=&\Gamma_3,\nonumber
\end{eqnarray}
which satisfy collectively (\ref{1a}) such that
$\Gamma_1+\Gamma_2+\Gamma_3=0$.

We define dimensionless density parameters via
\begin{equation}\label{3}
x\equiv\frac{\kappa^2\rho_d}{3H^2},\ \
y\equiv\frac{\kappa^2\rho_m}{3H^2},\ \
z\equiv\frac{\kappa^2\rho_u}{3H^2}.
\end{equation}
Making use of parameters in (\ref{3}) in the modified Friedmann
equation (\ref{1}) is
\begin{equation}\label{4}
(x+y+z)\Big( 1-\frac{3H^2}{\kappa^2}\frac{x+y+z}{\rho_c} \Big)=1.
\end{equation}
Using (\ref{1}) and (\ref{2}), we can write
\begin{equation}\label{5}
-\frac{\dot H}{H^2}=\frac{3}{2}(2-x-y-z)\Big(
1+\frac{w_dx+w_uz}{x+y+z} \Big).
\end{equation}
The equation of state parameter of the total fluid is
\begin{equation}\label{6}
w_{\text{tot}}=\frac{p}{\rho}=\frac{w_dx+w_uz}{x+y+z}.
\end{equation}
The continuity equations (\ref{2a}) in dimensionless variables
reduce to
\begin{eqnarray}\label{7}
x'&=&3x\Big( 1+\frac{w_dx+w_uz}{x+y+z}
\Big)(2-x-y-z)\nonumber\\&&-3x-3w_dx+\frac{\kappa^2}{3H^3}\Gamma_1,\nonumber\\
y'&=&3y\Big( 1+\frac{w_dx+w_uz}{x+y+z}
\Big)(2-x-y-z)\nonumber\\&&-3y+\frac{\kappa^2}{3H^3}\Gamma_2,\\
z'&=&3z\Big( 1+\frac{w_dx+w_uz}{x+y+z}
\Big)(2-x-y-z)\nonumber\\&&-3z-3w_uz+\frac{\kappa^2}{3H^3}\Gamma_3.\nonumber
\end{eqnarray}
Above the primes denote differentiation with respect to $N=\ln a$.
The coupling functions $\Gamma_i$, $i=1,2,3$ are in general
functions of the energy densities and the Hubble parameter i.e.
$\Gamma_i(H\rho_i)$. The system of equations in (\ref{7}) is
analyzed by first equating them to zero to obtain the critical
points. Next we perturb (\ref{7}) up to first order about the
critical points and check their stability.

Following \cite{2}, one can consider various coupling functions to
model the interaction. However all the models described here are
phenomenological. There is, at yet, no deep theoretical
justification for assuming this interaction in the absence of the
quantum gravity. There has been a striking resurgence in modeling
dark energy via such interacting models as they explain the
astrophysical data with the desired accuracy. Moreover, in such
models, one can obtain attractor solutions (stable equilibrium
points against perturbation) at which the energy densities come at
equilibrium and thereby explain cosmic coincidence. These models
have very little fine tuning problems since these contain only one
arbitrary coupling parameter. In literature, this parameter has been
constrained via astrophysical data  from supernovae, cosmic
background radiation, baryon acoustic oscillations, gas mass
fraction in galaxy clusters, the history of the Hubble function, and
the growth function \cite{die}. The signature of the coupling
parameter is of central importance: its positive (negative) sign
gives the direction of interaction.

There are some drawbacks as well for the interacting dark energy
models: they are not clearly distinguishable from the
non-interacting ones, in the light of the observational data
\cite{diego}. Another drawback is that there is arbitrariness in the
choice of the interaction term (i.e. products of Hubble parameter
with the energy densities); there are models in which only the
coupling parameter or the dark energy state parameter is taken as a
function of scale factor to model interaction without employing the
densities or the Hubble parameter \cite{li}.

\section{Analysis of stability in phase space}

In this section, we will construct four models by choosing different
coupling forms $\Gamma_i$ and analyze the stability of the
corresponding dynamical systems about the critical points. We shall
plot the phase and evolutionary diagrams accordingly.

\subsection{Interacting Model  I}

We consider the model with the following interaction terms
\begin{equation}\label{8a}
\Gamma_1=-6bH\rho_x, \ \ \Gamma_2=\Gamma_3=3bH\rho_x,
\end{equation}
where $b$ is a coupling parameter and we assume it to be a positive
real number of order unity. Thus (\ref{8a}) says that both matter
and unparticles have increase in energy density with time, while
dark energy loses its energy density. Therefore, it is a decay of
dark energy into matter and unparticle.

Using (\ref{8a}), the system (\ref{7}) takes the form
\begin{eqnarray}\label{8}
x'&=&3x\Big( 1+\frac{w_dx+w_uz}{x+y+z}
\Big)(2-x-y-z)\nonumber\\&&-3x-3w_dx-6bx,\nonumber\\
y'&=&3y\Big( 1+\frac{w_dx+w_uz}{x+y+z}
\Big)(2-x-y-z)\nonumber\\&&-3y+3bx,\\
z'&=&3z\Big( 1+\frac{w_dx+w_uz}{x+y+z}
\Big)(2-x-y-z)\nonumber\\&&-3z-3w_uz+3bx.\nonumber
\end{eqnarray}
There are three critical points:
\begin{itemize}
\item Point $A_1:\ (0,0,1)$,
\item Point $B_1:\ (0,1,0)$,
\item Point $C_1=\Big(
\frac{(w_d+2b)(w_d+2b-w_u)}{w_d(w_d+2b-w_u)+w_ub},\\
-\frac{b(w_d+2b-w_u)}{w_d(w_d+2b-w_u)-w_ub},
-\frac{b(w_d+2b)}{w_d(w_d+2b-w_u)-w_ub}\Big)$.
\end{itemize}
The eigenvalues of the Jacobian matrix for these critical points
are:
\begin{itemize}
\item Point $A_1:\lambda_1= w_u-w_d-2b, \lambda_2= w_u,\\
\lambda_3= -1-w_u,$
\item Point $B_1:\lambda_1= -(w_d+2b), \lambda_2=-1,
\lambda_3=-w_u$,
\item Point $C_1:\lambda_1=-(1+w_d+2b),\\ \lambda_2=-w_d\Big(\frac{(w_d+2b)(w_d+2b
-w_u)}{w_d(w_d+2b-w_u)+w_ub} \Big),\\ \lambda_3=
-\Big(\frac{(w_d+b)(w_d-w_u)}{w_d}+b\Big)$.
\end{itemize}
For the point $A_1$, the eigenvalue $\lambda_2$ is non-negative,
which indicates that $A_1$ is not a stable point. For the point
$B_1$, we also find that both of the eigenvalues $\lambda_1$ and
$\lambda_3$ are negative  when $w_d>-2b$. Therefore the point $B_1$
is  the stable point. For the point $C_1$ , when $w_d<-b$ and
$w_d<w_u$, all of the eigenvalues $\lambda_1$, $\lambda_2$ and
$\lambda_32$ are negative, which means that $C_1$ is a stable point.
As figure 1 shows below, the stable critical point $C_1$ gives us
the intuitive picture where the $x_c$ (representing DE) decreases
while $y_c$ (representing matter) and $z_c$ (representing
unparticle) increases. Moreover, the situation arises where matter
dominates asymptotically and unparticle density comes after matter.
Since $C_1$ is an attractor solution, it implies that the dynamical
equations (\ref{7}) yield this behavior for generic initial
conditions. It is interesting to note that similar situation arises
in Einstein's gravity as well \cite{2}. The stable region of $C_1$
is not affected by the EoS of the unparticle, but the position of
$C_1$ in the phase space is decided together by $w_d$, $w_u$ and the
coupling constant $b$. From (\ref{6}), we also learn that the
effective total EoS at point $C_1$ is
\begin{eqnarray}
w_{\text{tot}} &=&\Big[ {\frac {w_d ( 2\,b ( w_d+ 2\,b-w_u ) +w_d(
w_d+ 2\,b-w_u )  ) }{w_ d ( w_{{d}}+2\,b-w_u)
+bw_u}}\nonumber\\&&-{\frac {w_ub ( 2\,b+w_d ) }{w_{{d}} (
w_d+2\,b-w_u ) -bw_u}} \Big]\nonumber\\&&\times  \Big[ {\frac {2\,b
( w_{{d}}+ 2 \,b-w_{{u}} ) +w_{{d}} ( w_{{d}}+ 2\,b-w_{{u}} ) }{ w_d
( w_d+2\,b-w_u ) +bw_u}}\nonumber\\&&-{\frac {b (
w_{{d}}+2\,b-w_{{u}} ) }{w_{{d}} ( w_{{d}}+2\,b-w_{ {u}} )
-bw_u}}\nonumber\\&&-{\frac {b ( 2\,b+w_d ) }{w_d  (
w_{{d}}+2\,b-w_{{u}} ) -bw_u}} \Big] ^{-1}
\end{eqnarray}
The critical point $C_1$ denotes that the DE, DM and unparticle can
be coexisted in the late-times of the Universe.
\begin{figure}
\centering
 \includegraphics[scale=0.3] {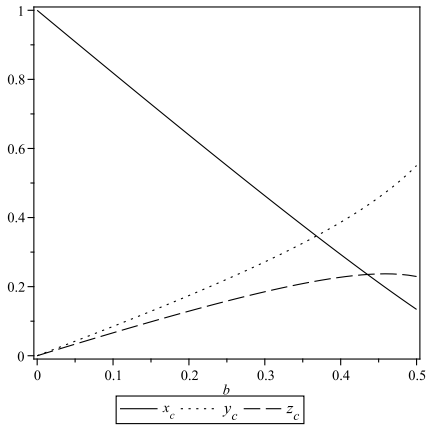}% scale goes from 0 to 1.
  \caption{ Variety of $x_c$, $y_c$ and $z_c$ with b at the critical point $C_1$ for fixed
   $w_d = -1.2$ and $w_u = 0.28$. Here the coupling
constant b is located in the region $w_d<-b$. }
  \label{1.eps}
\end{figure}
Figure (2) shows the evolution of the functions $(x,y,z)$ for a
variety set of parameters . As we observe, for a set of density
parameters $w_d=-1.2,w_u=0.28,b=0.5$, matter is the dominant portion
and the unparticle's density always remain below the dark energy.
One interesting feature is that for large values of scale factor
(the smaller values of the redshift $z$) the two different parts of
the matters in the Universe (unparticle and dark energy) tend to the
same value and after reaching to this point, the dominant part is
the matter field.

\begin{figure}
\centering
 \includegraphics[scale=0.3] {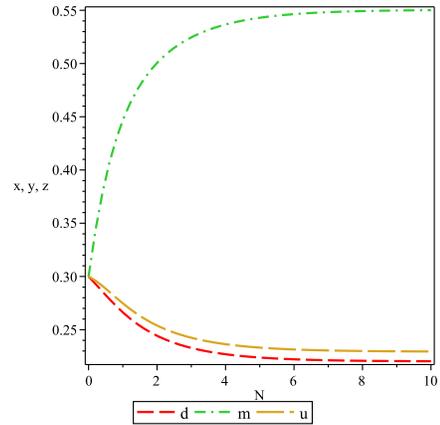}% scale goes from 0 to 1.
  \caption{ Variety of $x,y,z$ as a function of the $N=\ln(a)$. The initial
  conditions chosen are $x(0)=y(0)=z(0)=0.3$, $w_u=0.28$ and
  $w_d=-1.2$ and $b=0.5$. We observe that the dominant part
  of in Model I is matter. }
  \label{2-1.eps}
\end{figure}
Now we change the parameters to a new set $w_d=-1.7$, $w_u=0.35$,
$b=0.25$. The behavior of the functions (densities) are very
different. Now, if the evolution begins from a very large negative
red shift, the dominant part of the model for all times is the dark
energy. As the previous case, the unparticle's portion remains under
dark energy and matter. For some values of large scale factor, the
three different parts of the matters fields (dark energy, matter,
unparticle) reach to the same asymptotic's value. These evolutionary
scheme has been shown in figure (3)
\begin{figure}
\centering
 \includegraphics[scale=0.3] {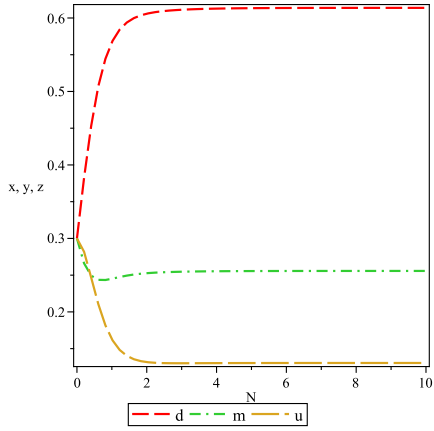}% scale goes from 0 to 1.
  \caption{ Variety of $x,y,z$ as a function of the $N=\ln(a)$. As we observe, the dominant part
  of the model is dark energy. The initial
  conditions chosen are $x(0)=y(0)=z(0)=0.3$, $w_u=0.28$ and
  $w_d=-1.2$ and $b=0.5$.}
  \label{2-2.eps}
\end{figure}
 As we observe in figure (4), for a set of density
parameters $-1<w_d<-\frac{1}{3},w_u>0,b=0.5$, $w_{tot}>0$, matter is
the dominant portion and the unparticle's density always remain
below the dark energy. It is consistent with the result shown in
figure (2) in which as the coupling is strong enough the Universe
will be dominated by DM.
\begin{figure}
\centering
 \includegraphics[scale=0.35] {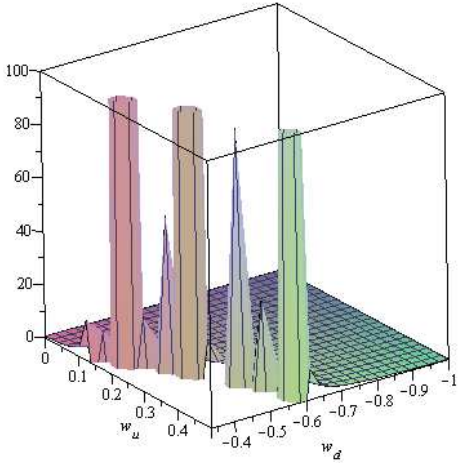}% scale goes from 0 to 1.
  \caption{ Variety of $w_{tot}$ as a function of the $w_d,w_u$. AS we observe,
  $w_{\text{tot}}>0$. }
  \label{6.eps}
\end{figure}
Figure (5) shows the  phase diagram of interacting dark energy with
DM and unparticle in loop quantum cosmology through the coupling
terms. The point $C_1$ is the critical point. Here we choose the
values $w_d=-1.7,w_u=0.28,b=0.5$ in the stable region
 $w_d<-b$ and $w_d<w_u$.
\begin{figure}
\centering
 \includegraphics[scale=0.35] {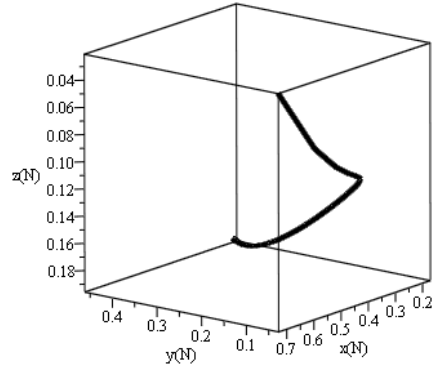}% scale goes from 0 to 1.
  \caption{ The  phase diagram of interacting dark energy
with DM and unparticle in loop quantum cosmology through the
coupling terms. Here we choose the values $w_d=-1.7,w_u=0.28,b=0.5$
in the stable region
 $w_d<-b$ and $w_d<w_u$. }
  \label{7.eps}
\end{figure}

\subsection{Interacting Model II}

We study another model with the choice of the interaction terms
\begin{equation}\label{9a}
\Gamma_1=-3bH\rho_d,\ \ \Gamma_2=3bH(\rho_d-\rho_m),\ \
\Gamma_3=3bH\rho_m.
\end{equation}
This model effectively describes the situation when dark energy
loses energy density to matter while the unparticle density
increases due to interaction with the matter.

\begin{eqnarray}\label{9}
x'&=&3x\Big( 1+\frac{w_dx+w_uz}{x+y+z}
\Big)(2-x-y-z)\nonumber\\&&-3x-3w_dx-3bx,\nonumber\\
y'&=&3y\Big( 1+\frac{w_dx+w_uz}{x+y+z}
\Big)(2-x-y-z)\nonumber\\&&-3y+3b(x-y),\\
z'&=&3z\Big( 1+\frac{w_dx+w_uz}{x+y+z}
\Big)(2-x-y-z)\nonumber\\&&-3z-3w_uz+3by.\nonumber
\end{eqnarray}

There are three critical points:
\begin{itemize}
\item Point $A_2:\ (0,0,1)$,
\item Point $B_2:\ (0,1-\frac{b}{w_u},\frac{b}{w_u})$,
\item Point $C_2=\Big( \frac{w_d(b+w_d-w_u)}{w_d(w_d-w_u)+bw_u}
,\\-\frac{b(w_d-b-w_u)}{w_d(w_d-w_u)+bw_u}
,\frac{b^2}{w_d(w_d-w_u)+bw_u}\Big)$.
\end{itemize}

The eigenvalues of the Jacobian matrix for these critical points
are:
\begin{itemize}
\item Point $A_2:\lambda_1= w_u-w_d-b, \lambda_2= w_u-b,\\
\lambda_3= -(1+w_u),$
\item Point $B_2:\lambda_1= -w_d, \lambda_2=-(1+b),\\
\lambda_3=\frac{b}{w_u}(w_u-1-2b)+b$,
\item Point $C_2:\lambda_1=-(1+w_d+b),\\ \lambda_2=-w_d\Big(\frac{w_d(b+w_d-w_u)}{w_d(w_d-w_u)
+bw_u}\Big),\\ \lambda_3= -\Big(\frac{(w_d-b)(w_d-w_u)}{w_d}\Big)$.
\end{itemize}

\begin{figure}
\centering
 \includegraphics[scale=0.3] {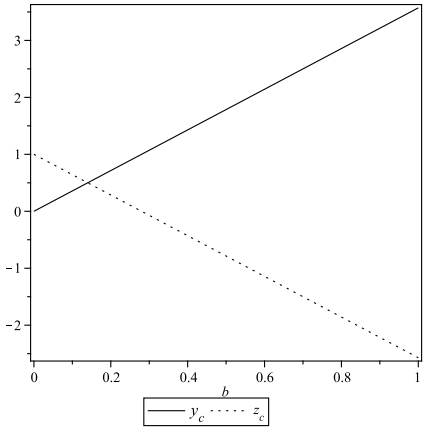}% scale goes from 0 to 1.
  \caption{ Variety of $x_c$, $y_c$ and $z_c$ with b at the critical point $B_2$ for fixed
   $w_d = -1.2$ and $w_u = 0.28$.  }
  \label{3.eps}
\end{figure}
For point $A_2$, the eigenvalue $\lambda_i<0$ if $b>w_u-w_d$ for
$w_d<0$, hence point $A_2$ is a stable critical point. The second
critical point $B_2$ is  unstable since $\lambda_1>0$. It is easy to
see that the third critical point $C_2$ is stable when we have the
next set of inequalities
\begin{eqnarray}
-(1+w_d)<b<w_d\\
w_u-w_d<b,b<\frac{w_d}{w_u}(w_u-w_d)
\end{eqnarray}
Figure (3) shows that all the three components start with equal
energy densities, they evolve differently. Here the dark energy
dominates over matter and unparticle at late times. The energy
density of dark energy evolves to $x\sim0.64$ and $y\sim0.23$
compatible with the observations, while the unparticle density falls
to zero at an early epoch $N\sim2$. Clearly this phenomenological
model of dynamical interaction explains the present state of the
Universe.
\begin{figure}
\centering
 \includegraphics[scale=0.3] {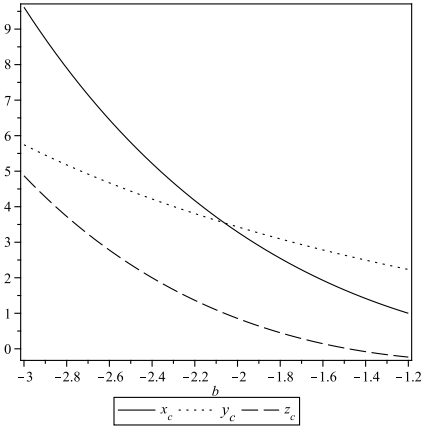}% scale goes from 0 to 1.
  \caption{ Variety of $x_c$, $y_c$ and $z_c$ with b at the critical point $C_2$ for fixed
   $w_d = -1.2$ and $w_u = 0.28$. Here the coupling
constant b is located in the region $b<w_d,w_d<w_u$. }
  \label{4.eps}
\end{figure}
The next figure (8), shows, the  phase diagram of interacting dark
energy with DM and unparticle in loop quantum cosmology through the
coupling terms . The point $C_2$ is the stable critical point. Here
we choose the values $w_d=-1.7,w_u=0.28,b=0.5$ in the stable region
$ -(1+w_d)<b<w_d, w_u-w_d<b,b<\frac{w_d}{w_u}(w_u-w_d)$.
\begin{figure}
\centering
 \includegraphics[scale=0.35] {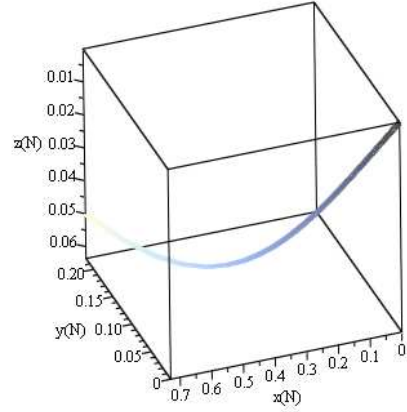}% scale goes from 0 to 1.
  \caption{ The  phase diagram of interacting dark energy
with DM and unparticle in loop quantum cosmology through the
coupling terms. Here we choose the values $w_d=-1.7,w_u=0.28,b=0.5$
in the stable region
 $
-(1+w_d)<b<w_d, w_u-w_d<b,b<\frac{w_d}{w_u}(w_u-w_d)$.  }
  \label{8.eps}
\end{figure}

\subsection{Interacting Model - III}

The coupling terms containing the product of energy densities have
been studied previously with the intent to reveal some new behavior
of the dynamical systems \cite{6}. Let us take the interaction terms
\cite{2}
\begin{equation}\label{12}
\Gamma_1=-6b\kappa^2 H^{-1}\rho_d\rho_u, \ \
\Gamma_2=\Gamma_3=3b\kappa^2 H^{-1}\rho_d\rho_u.
\end{equation}
The system in (\ref{7}) takes the form
\begin{eqnarray}
x'&=&3x\Big( 1+\frac{w_dx+w_uz}{x+y+z}
\Big)(2-x-y-z)\nonumber\\&&-3x-3w_dx-18bxz,\nonumber\\
y'&=&3y\Big( 1+\frac{w_dx+w_uz}{x+y+z}
\Big)(2-x-y-z)\nonumber\\&&-3y+9bxz,\\
z'&=&3z\Big( 1+\frac{w_dx+w_uz}{x+y+z}
\Big)(2-x-y-z)\nonumber\\&&-3z-3w_uz+9bxz.\nonumber
\end{eqnarray}
There are five critical points:
\begin{itemize}
\item Point $A_3:\ (1,0,0)$,
\item Point $B_3:\ (0,1,0)$,
\item Point $C_3:\ (0,0,1) $,
\item Point $D_3: \Big(\frac{1}{3}\frac{(1+w_u)}{b}, -\frac{1}{6}\frac{1+w_u+w_d+w_d w_u}{b},
 -\frac{1}{6}\frac{1+w_d}{b} \Big)$  ,
\item Point  $E_3: \Big(\frac{1}{3}\frac{w_u(w_d-w_u+6b)}{b(2w_d-w_u+6b)},\\  \frac{1}{3}
\frac{-9w_ub+w_u^2+18b^2+9w_d b-2w_dw_u+w_u^2}{b(2w_d-w_u+6b)},\\
-\frac{1}{3}\frac{w_d(w_d-w_u+3b)}{b(2w_d-w_u+6b)}\Big).$
\end{itemize}
\begin{figure}
\centering
 \includegraphics[scale=0.3] {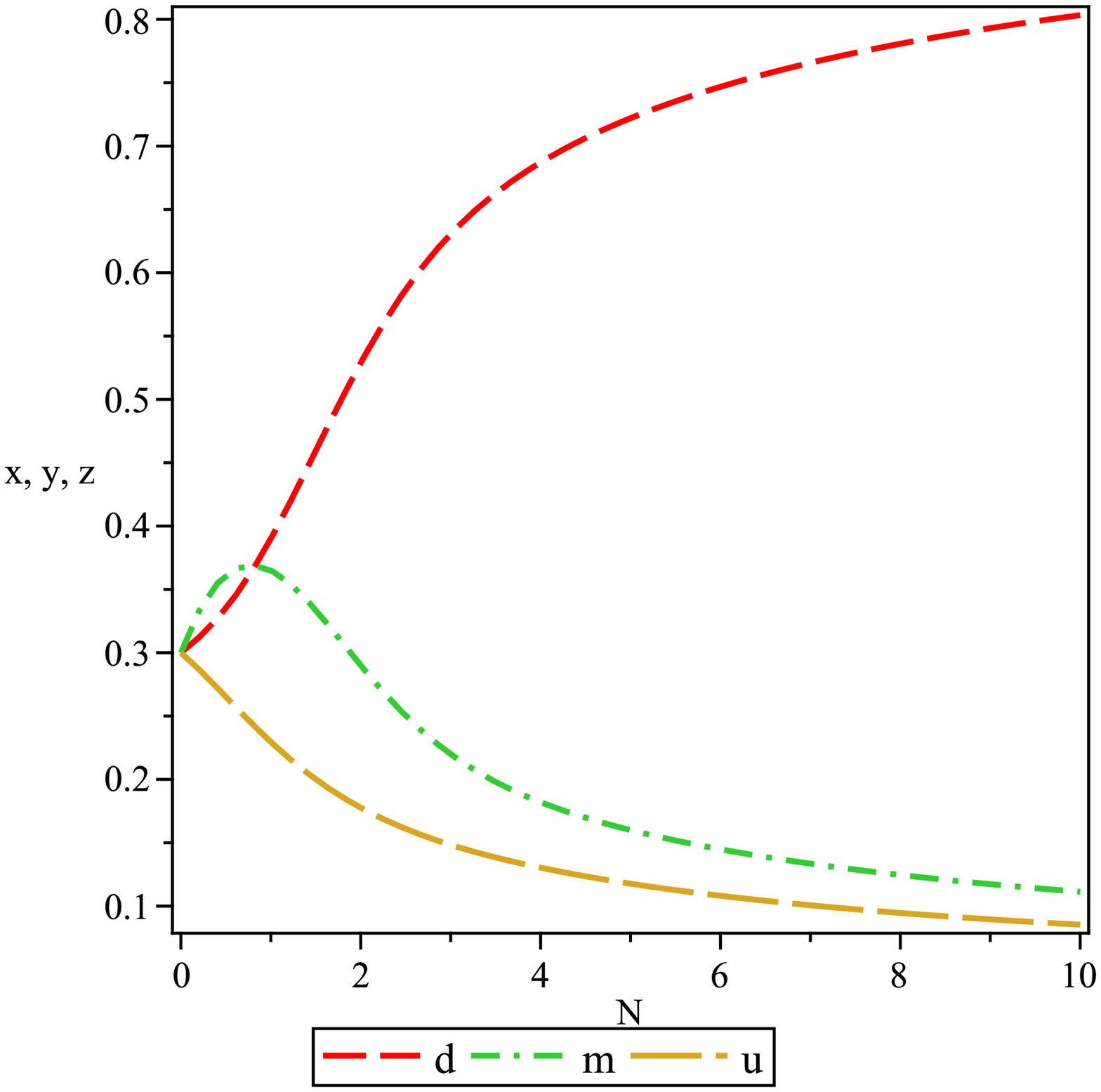}% scale goes from 0 to 1.
  \caption{ Variety of $x,y,z$ as a function of the $N=\ln(a)$. The initial
  conditions chosen are $x(0)=y(0)=z(0)=0.3$, $w_u=0.28$,
  $w_d=-1.2$ and $b=0.5$. }
\end{figure}
Now we must diagonalize the Jacobian matrix near these critical
points. For points $A_3,B_3,C_3$ we have
\begin{itemize}
\item Point $A_3:\ \lambda_1=-3(1+w_d),\lambda_2=3w_d,\lambda_3=3(3b-w_u+w_d)$,
\item Point $B_3:\ \lambda_1=-3,\lambda_2=-3w_u,\lambda_3=-3w_d$,
\item Point $C_3:\ \lambda_1= -3(1+w_u),\lambda_2=3w_u,\lambda_3=3(-6b+w_u-w_d)$,
\end{itemize}
For point $A _3$ since always $w_u>0,w_d\leq-1$ thus this point is
un stable. Similarly, the point $B_3,C_3$ both are unstable. The
analysis of stability for points $D_3,E_3$ are so complicated.
Indeed , the Jacobian matric in these cases, are not diagonal and
the behaviors of the eigenvalues are not trivial. Theoretically, we
cannot distinguish between these points as the stable or un stable
points. Thus, it is computationally and analytically impossible to
analyze these cases. Using the same initial conditions,Figure (9)
shows that dark energy density rises while matter density dominates
over dark energy and unparticle till $N\sim1$. For $N>1$, dark
energy density rises indefinitely (behaving like phantom energy)
while matter and unparticle density tends to zero.

\subsection{Interacting Model - IV}

Consider another model with the interaction terms \cite{2}
\begin{eqnarray}\label{11}
\Gamma_1&=&-3b\kappa^2 H^{-1}\rho_x\rho_u,\nonumber\\
\Gamma_2&=&3b\kappa^2 H^{-1}(\rho_x\rho_u-\rho_m\rho_u),\nonumber\\
\Gamma_3&=&3b\kappa^2 H^{-1}\rho_m\rho_u.
\end{eqnarray}
The system in (\ref{7}) takes the form
\begin{eqnarray}
x'&=&3x\Big( 1+\frac{w_dx+w_uz}{x+y+z}
\Big)(2-x-y-z)\nonumber\\&&-3x-3w_dx-9bxz,\nonumber\\
y'&=&3y\Big( 1+\frac{w_dx+w_uz}{x+y+z}
\Big)(2-x-y-z)\nonumber\\&&-3y+9b(xz-yz),\\
z'&=&3z\Big( 1+\frac{w_dx+w_uz}{x+y+z}
\Big)(2-x-y-z)\nonumber\\&&-3z-3w_uz+9byz.\nonumber
\end{eqnarray}
\begin{figure}
\centering
 \includegraphics[scale=0.3] {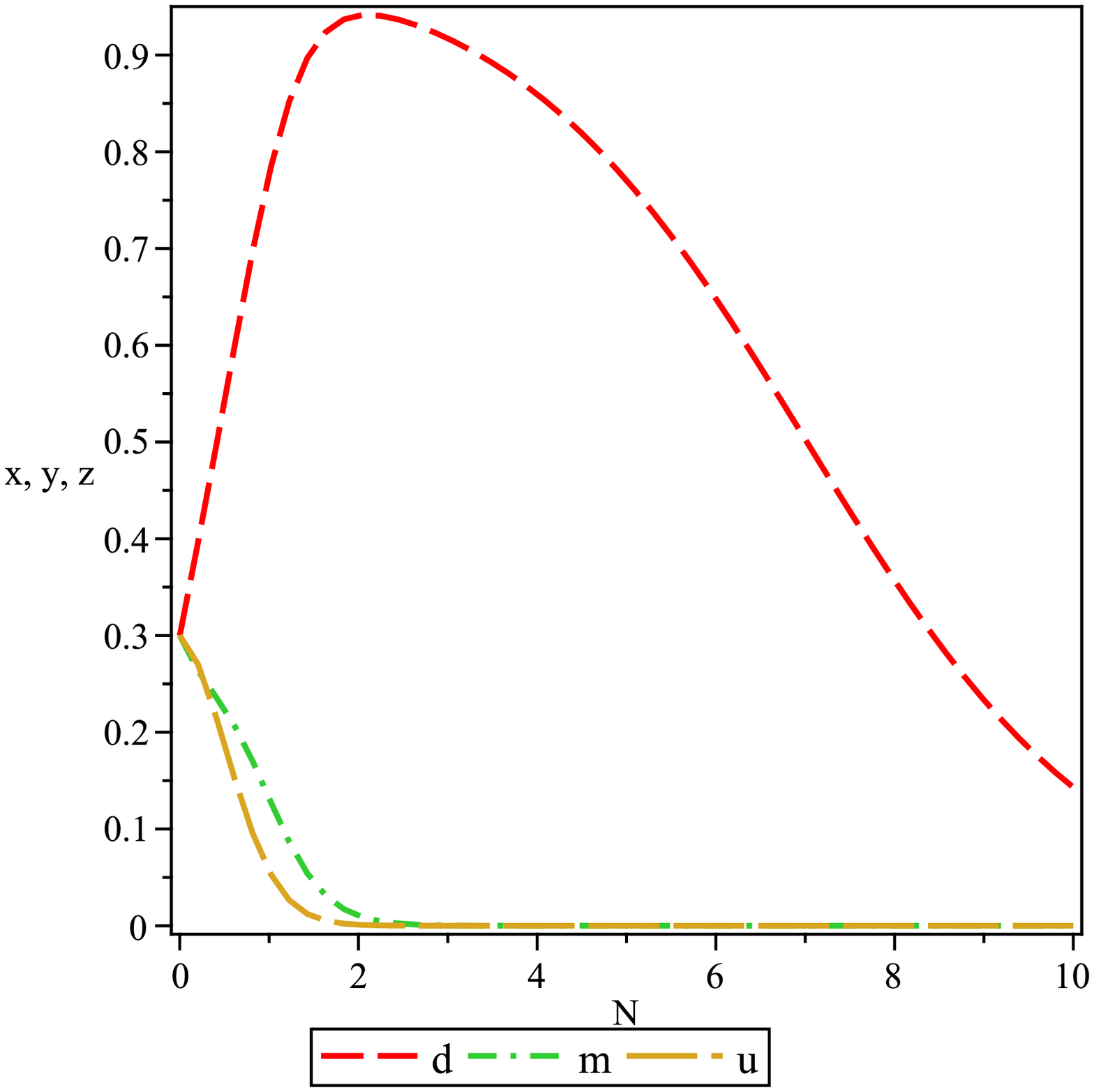}% scale goes from 0 to 1.
  \caption{ Variety of $x,y,z$ as a function of the $N=\ln(a)$. The initial
  conditions chosen are $x(0)=y(0)=z(0)=0.3$, $w_u=0.28$,
  $w_d=-1.2$ and $b=0.5$. }
\end{figure}
There are six critical points:
\begin{itemize}
\item Point $A_4:\ (1,0,0)$,
\item Point $B_4:\ (0,1,0)$,
\item Point $C_4:\ (0,0,1) $,
\item Point $D_4: \Big(0,\frac{1}{3}\frac{(1+w_u)}{b},-\frac{1}{3b}
\Big)$,
\item Point  $E_4: \Big(\frac{1}{3}\,{\frac
{-w_{{u}}+3b+w_{{d}}}{b}},\\
\frac{1}{3}\,{\frac{-2w_{{d}}w_{{u}}+{w_{{u}}}^{2}+3w_{{d}}b+{w_{{d}}}^{
2}-3w_{{u}}b}{b ( -w_{{u}}+3\,b ) }},  \frac{1}{3}\,{\frac {w_{{d}}
( w_{{d}}-w_{{u}} ) }{b ( -w_{{u} }+3b ) }} \Big )$,
\item Point  $F_4:  \Big (\frac{1}{3}\,{\frac {w_{{d}} ( 1+w_{{u}} ) }{b ( 1+w_{{d}}
) }} ,\frac{1}{3}\,{\frac {1+w_{{u}}}{b}},-\frac{1}{3}\,{\frac
{1+w_{{d}}}{b}} \Big).$
\end{itemize}
Now we must diagonalize the Jacobian matrix near these critical
points. For points $A_4,B_4,C_4$ we have
\begin{itemize}
\item Point $A_4:\ \lambda_1=-3(1+w_d),\lambda_2=3w_d,\\ \lambda_3=3(w_d-w_u)$,
\item Point $B_4:\ \lambda_1=-3,\lambda_2=-3w_d,\lambda_3=3(3b-w_u)$,
\item Point $C_4:\ \lambda_1= -3(1+w_u),\lambda_2=3(w_u-3b),\\
\lambda_3=3(-3b+w_u-w_d)$.
\end{itemize}
For point $A _4$ since always $w_u>0,w_d\leq-1$ thus this point is
 un stable, specially for the cross line $w_d=-1$ which in this case, the system is
  unstable near $A_4$.
 Similarly, the point $B_4$ is unstable. But the point $C_4$ is a stable point if
$w_d<-1,$ and $b>(w_u-w_d)/3$. The analyses of stability for points
$D_4$ and $E_4$ are as complicated as the Model III. Indeed, the
Jacobian matric in these cases, are not diagonal and the behaviors
of the eigenvalues are not trivial. Theoretically, we cannot
distinguish between these points as stable or unstable points. Thus,
it is computationally and analytically impossible to analyze these
cases. However, the behavior of the dynamical equations in (20) is
plotted in Figure (10) which shows that dark energy density rises
rapidly till $N\sim2.5$ after it decreases sharply (behaving like
quintessence), while matter and unparticle density decrease and
approach to zero at $N\geq2$. It is clear that dark energy does not
decay into matter or unparticle, otherwise their densities would
have increased. Hence the decay of dark energy into some mysterious
component is not clear.

\section{Discussion}

In this paper, we have studied the dynamical behaviors when dark
energy is coupled to dark matter and the unparticle in the
background of flat FRW spacetime. The analysis was performed in the
theoretical framework of loop quantum cosmology which is one of the
emerging fields of quantum cosmology. It was assumed that the
correction terms due to LQC remain benign and become effective near
the Planck density. To address the coincidence problem, we
introduced a phenomenological interaction between dark energy, dark
matter and unparticle. We constructed four toy models: here Model I
describes a stable attractor solution for a dark matter dominated
Universe; Model II also contains a stable attractor solution for a
dark energy dominated Universe; Model III and IV are dynamically
unstable since either they give completely dark energy filled
Universe or completely devoid of it.

Comparison of our LQC dynamical models with those of Einstein
cosmology \cite{2} reveals the following differences: the form of
most of the critical points (and hence eigenvalues) are different,
consequently the conditions and regions of stability of these
critical points also differs from Einstein cosmology. Also a
comparison of Fig. 1 above with Fig.2 of \cite{2} shows that the
evolution of energy densities is much slower in LQC than in
Einstein's case, however, the asymptotic evolution is the same.
Finally, unlike \cite{2} where all the considered models were
stable, we have only two stable scaling solutions, namely model I
and II. In other words, LQC results are very different from
Einstein's relativistic cosmology.

\section*{Acknowledgement}
The authors would like to thank the anonymous referees  for helpful
comments and suggestions. Also thanks are due to Ahmad Sheykhi for
useful communications related to this work.

\end{document}